\documentclass[11pt]{article}
\usepackage{graphicx}
\usepackage{dcolumn}
\usepackage{bm}

\usepackage{epsfig}

\newcommand{\gp}{$\vec \gamma p \rightarrow p\pi^+\pi^-${ }}

\begin{document}

\begin{center}
\renewcommand{\baselinestretch}{1.2}\normalsize
{\Large\bf Helicity-Dependent Angular Distributions 
           in Double-Charged-Pion Photoproduction}

\renewcommand{\baselinestretch}{1.5}\normalsize
S.~Strauch for the CLAS Collaboration

{\it The George Washington University, Washington, D.C. 20052, USA}  
\today
\vspace*{0.5cm}
\end{center}

\begin{abstract}
Two-pion photoproduction in the reaction \gp has been studied at Jefferson
Lab Hall B using a circularly-polarized tagged photon beam in the
energy range between 0.6~GeV and 2.3~GeV.  Owing to the large angular
acceptance of the CLAS detector, complete beam-helicity-dependent
angular distributions of the final-state particles were
measured. The large cross-section asymmetries exhibit strong
sensitivity to the kinematics of the reaction and provide valuable
information on the reaction dynamics. Preliminary results are
presented.
\end{abstract}

\section{Introduction}

Many nucleon resonances in the mass region above 1.6~GeV decay
predominantly through $\Delta\pi$ or $N\rho$ intermediate states into
$N\pi\pi$ final states (see the Particle-Data Group review,
\cite{pdg02}). This makes electromagnetic exclusive double-pion
production an important tool in the investigation of $N^*$ structure
and reaction dynamics, as well as in the search for ``missing''
baryon states. Unpolarized cross-section measurements of double-pion
electroproduction have recently been reported in
\cite{Ripani03}. Further constraints are to be found in polarization
observables.

Here, for the first time, a measurement of the \gp reaction is being
reported, where the photon beam is circularly polarized and no
nuclear polarizations (target or recoil) are specified. The general
form of the cross section can then be written as%
\footnote{See \protect\cite{Donnelly96} for an overview of the more
general case of exclusive-$n$ electron scattering, $A(e,e'x_1\ldots
x_n)$. The real-photon case, discussed here, results in a simpler
formalism.}

\begin{equation}
   \sigma = \Sigma + P_\odot\Delta\; ,
\end{equation}

\noindent where $P_\odot$ is the degree of circular polarization of
the photon and $\Sigma$ and $\Delta$ are the unpolarized and polarized
cross sections, respectively. For this kind of study, a three-body
final state is crucial, since reactions with two-body final states are
always coplanar and have identical cross sections for unpolarized or
circularly polarized photons \cite{Boffi96}, so that $\Delta =
0$. 

The cross-section asymmetry, obtained by flipping the beam polarization,
is defined by:
\begin{equation}
   A = \frac{1}{P_\odot} \cdot 
   \frac{\sigma^+ - \sigma^-}{\sigma^+ + \sigma^-} = \frac{\Delta}{\Sigma} \;.
\end{equation}
In this paper, we will discuss the double-charged-pion channel from a
hydrogen target.

\section{Data}

The \gp reaction was studied with the CEBAF Large Acceptance
Spectrometer (CLAS) \cite{Mecking03} at Jefferson Lab.  Longitudinally
polarized electrons with an energy of 2.4~GeV were incident on the
thin radiator of the Hall-B Photon Tagger \cite{Sober00} and produced
circularly-polarized tagged photons in the energy range between
0.6~GeV and 2.3~GeV. The collimated photon beam irradiated a
liquid-hydrogen target. The circular polarization of the photon beam
can be determined from the electron beam polarization and the ratio of
photon and incident electron energy \cite{Maximon59}.

To identify the reaction channel, the missing-mass technique was used,
requiring the detection of at least two out of three final-state
particles ($p$, $\pi^+$, and $\pi^-$) in the CLAS.
\begin{figure}[h!]
\centerline{
\epsfig{file=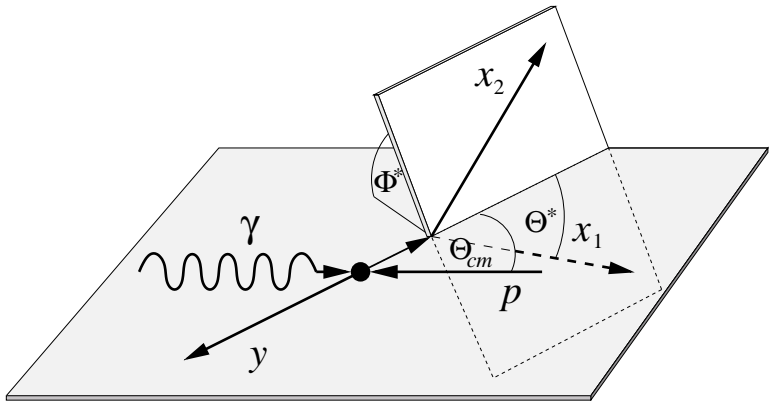,width=4.0in}
}
\caption{Angle definitions for the circular polarized real-photon
reaction $p(\vec \gamma,x_1 x_2)y$; $\Theta_{cm}$ is defined in the
center-of-mass frame of the $\gamma p$ system, $\Theta^*$ is defined
in the center-of-mass frame of the $x_1 x_2$ system.}
\label{fig:def}
\end{figure}
Owing to the large angular acceptance of the CLAS, complete angular
distributions of the cross-section asymmetries were observable. Figure
\ref{fig:def} defines the azimuthal angle $\Phi^*$ between the
scattering plane, containing the photon and the recoiling particle
$y$, and the decay plane, containing the two particles $x_1$ and
$x_2$.  There are three different configurations relevant for the \gp
reaction channel: $y[x_1,x_2]= p[\pi^+,\pi^-],\pi^+[p,\pi^-],$ and
$\pi^-[p,\pi^+]$.

\section{Results}

\begin{figure}[h!]
\centerline{\epsfig{file=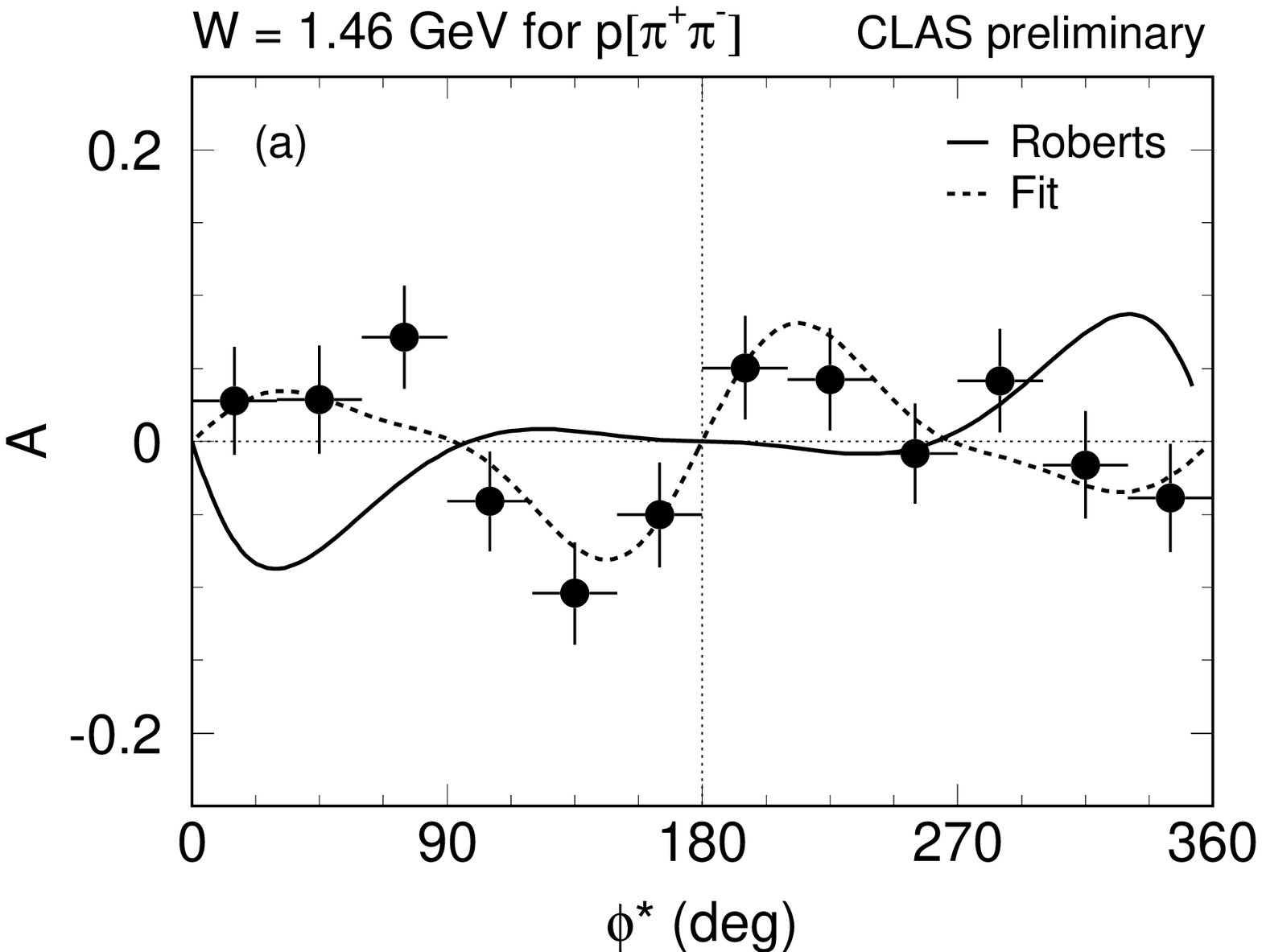,width=2.7in}%
            \epsfig{file=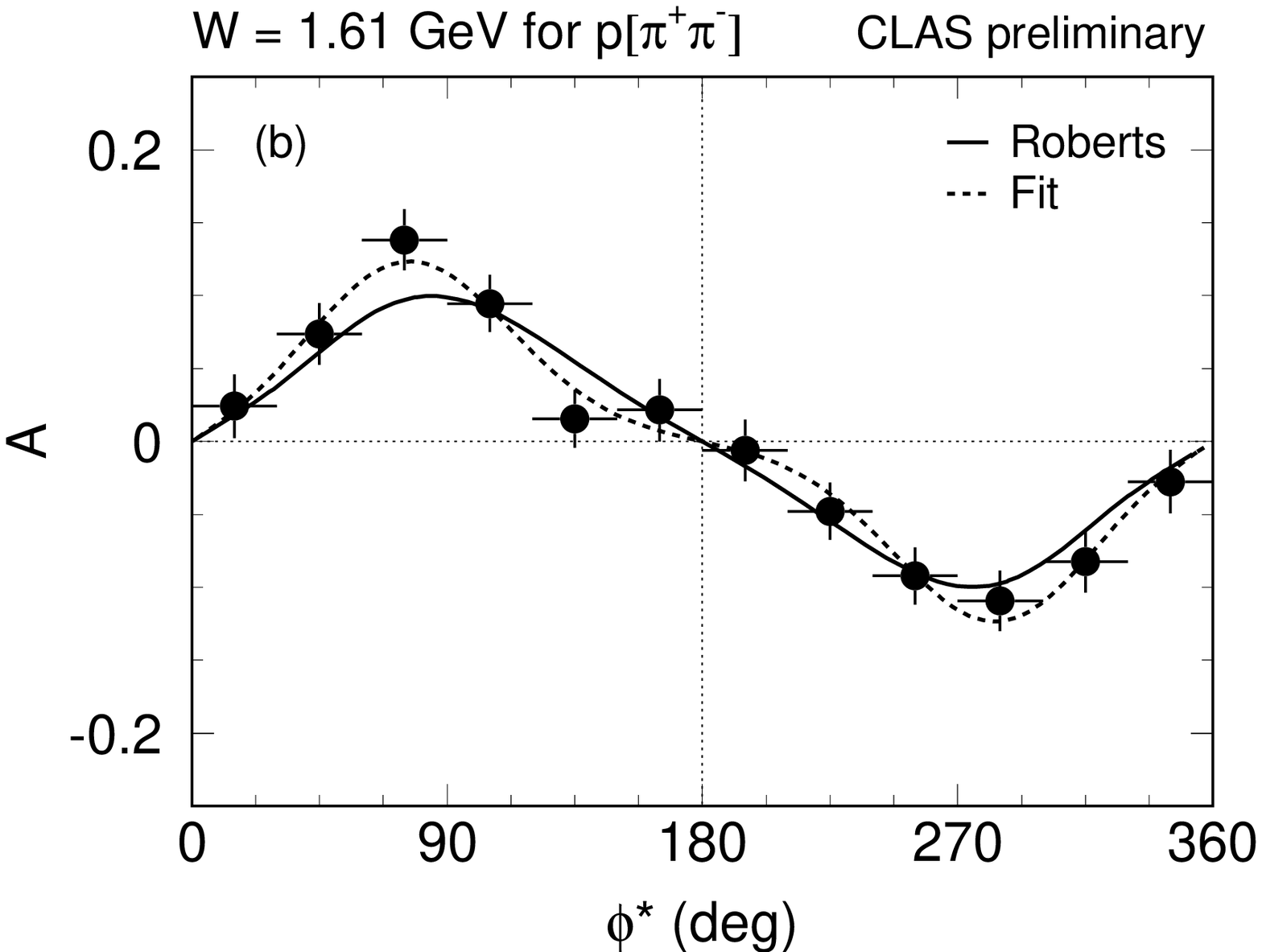,width=2.7in}}
\centerline{\epsfig{file=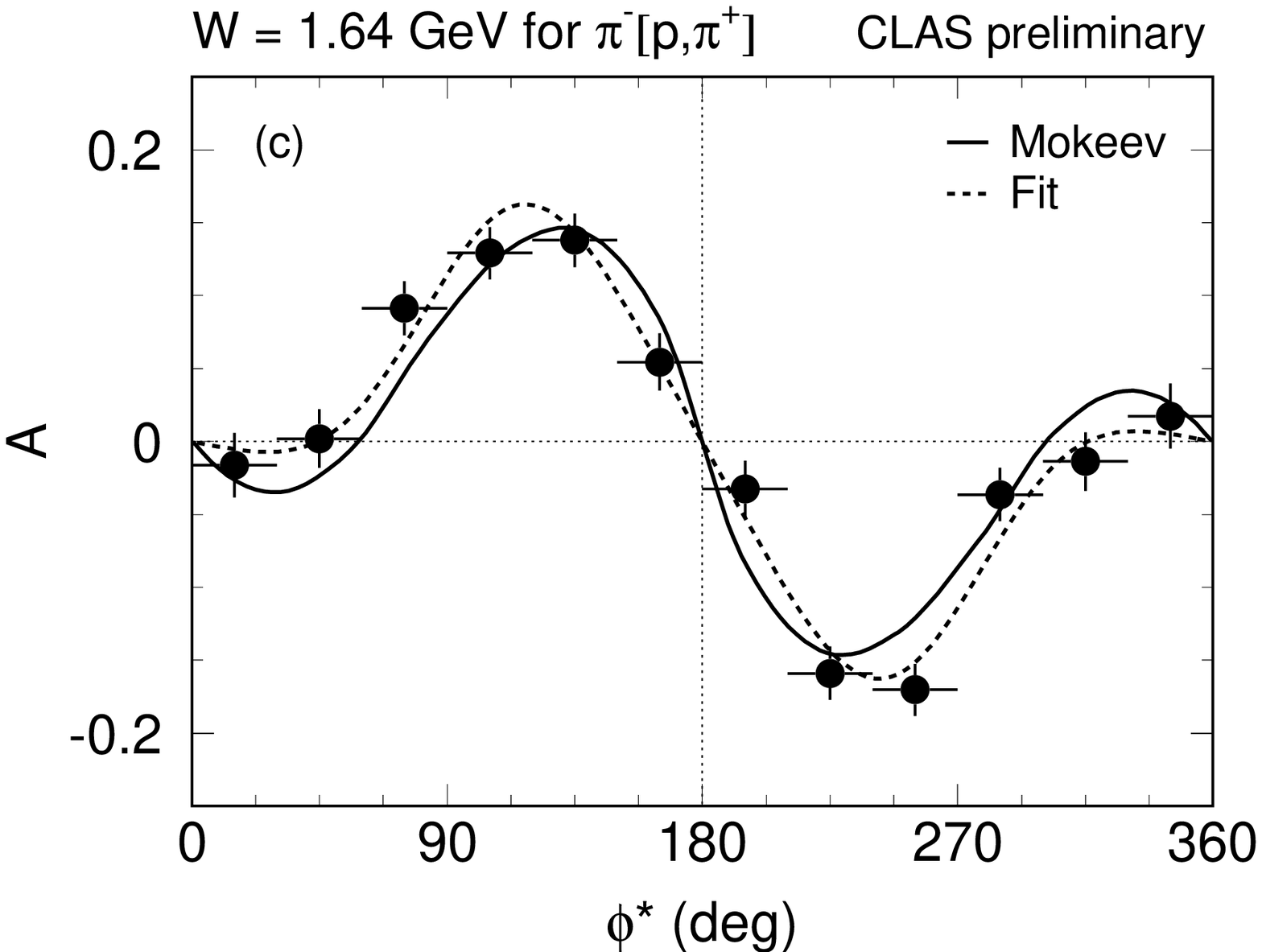,width=2.7in}%
            \epsfig{file=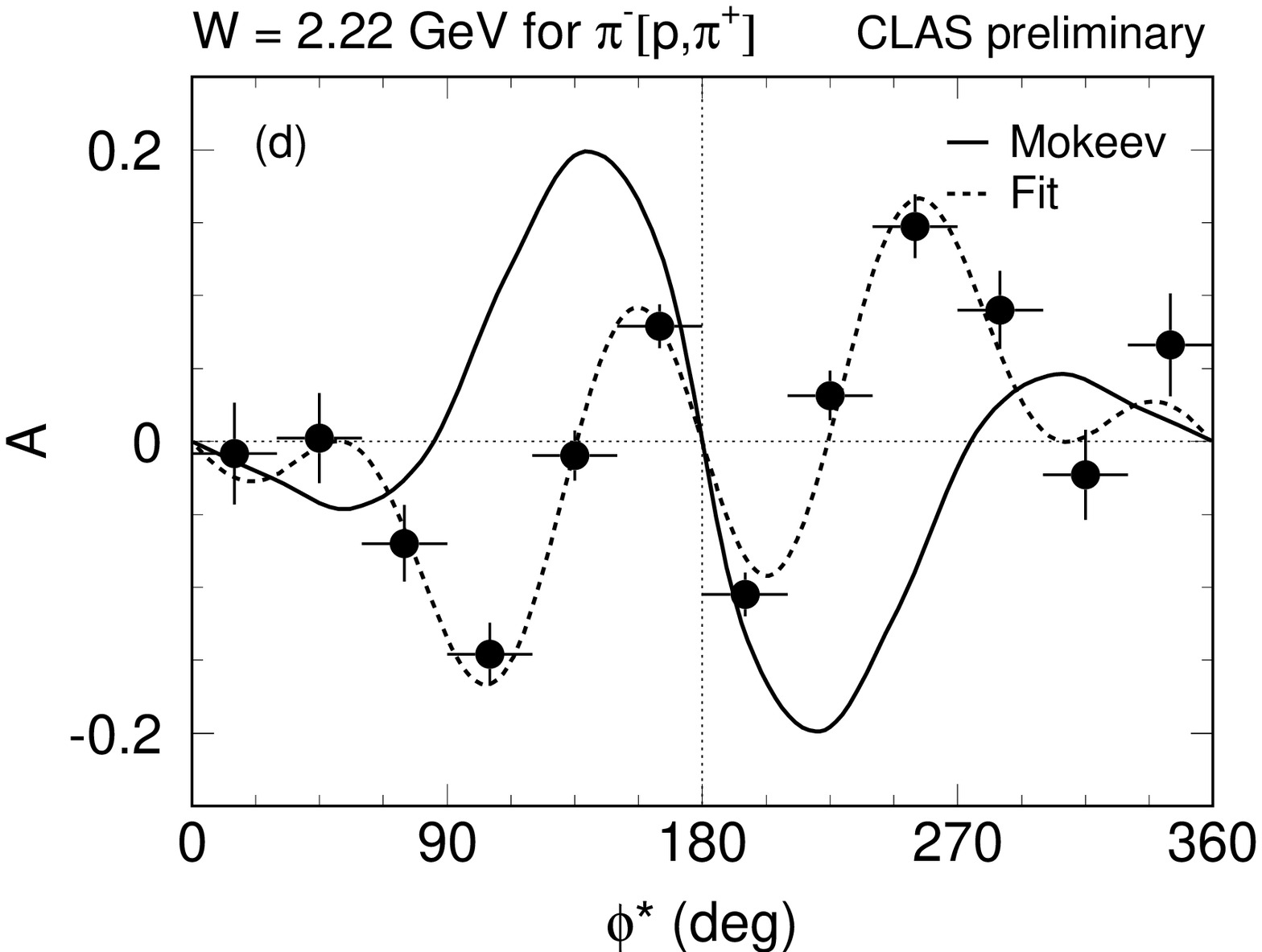,width=2.7in}}
\caption{Preliminary angular distribution for four different
center-of-mass energy bins ($\Delta W \approx 15$ MeV) of the
cross-section asymmetry in the \gp reaction; (a) and (b) for the
configuration $\pi^-[p,\pi^+]$, (c) and (d) for the configuration
$p[\pi^+,\pi^-]$.  The solid curves are calculations by Roberts
\protect\cite{Oed03,Roberts97} and Mokeev \protect\cite{Mokeev01} as
noted, the dotted curve results from fitting the data with
Eq.~(\ref{eq:fit}) up to order $n=4$.}
\label{fig:phi}
\end{figure}

Preliminary analysis of the double-charged-pion channels has revealed
large asymmetries in the helicity-dependent cross section.  Figure
\ref{fig:phi} shows preliminary $\Phi^*$ angular distributions for
different energy bins and for the $p[\pi^+,\pi^-]$ and
$\pi^-[p,\pi^+]$ configurations.  The asymmetry was fitted with a
series of sine functions:
\begin{equation}
A(\Phi^*) = \sum_{i=1}^{n} a_i \cdot \sin(i\Phi^*) \; .
\label{eq:fit}
\end{equation}
No cosine terms were included in this series because the cosine is an
even function. For example, when $\Phi^*=0$ or $180^\circ$, all the
particles are in one plane, and for this case, there is no helicity
dependence of the cross section, due to parity conservation. The fit
of Eq.~(\ref{eq:fit}) to the data is shown in Fig.~\ref{fig:phi} as
the dotted curves. Preliminary calculations for these cross-section
asymmetries were done by Oed and Roberts \cite{Oed03,Roberts97} using
a phenomenological Lagrangian approach and by Mokeev \cite{Mokeev01}
in a phenomenological calculation using available information on $N^*$
and $\Delta$ states. It is important to note that the calculations
performed to date have been integrated over $4\pi$, whereas the
experimental data are integrated only over the CLAS acceptance. The
results of the calculations are shown in Fig.~\ref{fig:phi} as solid
curves. At about $W = 1.6$~GeV, an excellent description of the data
is achieved by both calculations, whereas for the other energy bins
there is clearly room for improvement in the model parameters.

\begin{figure}[p]
\centerline{\epsfig{file=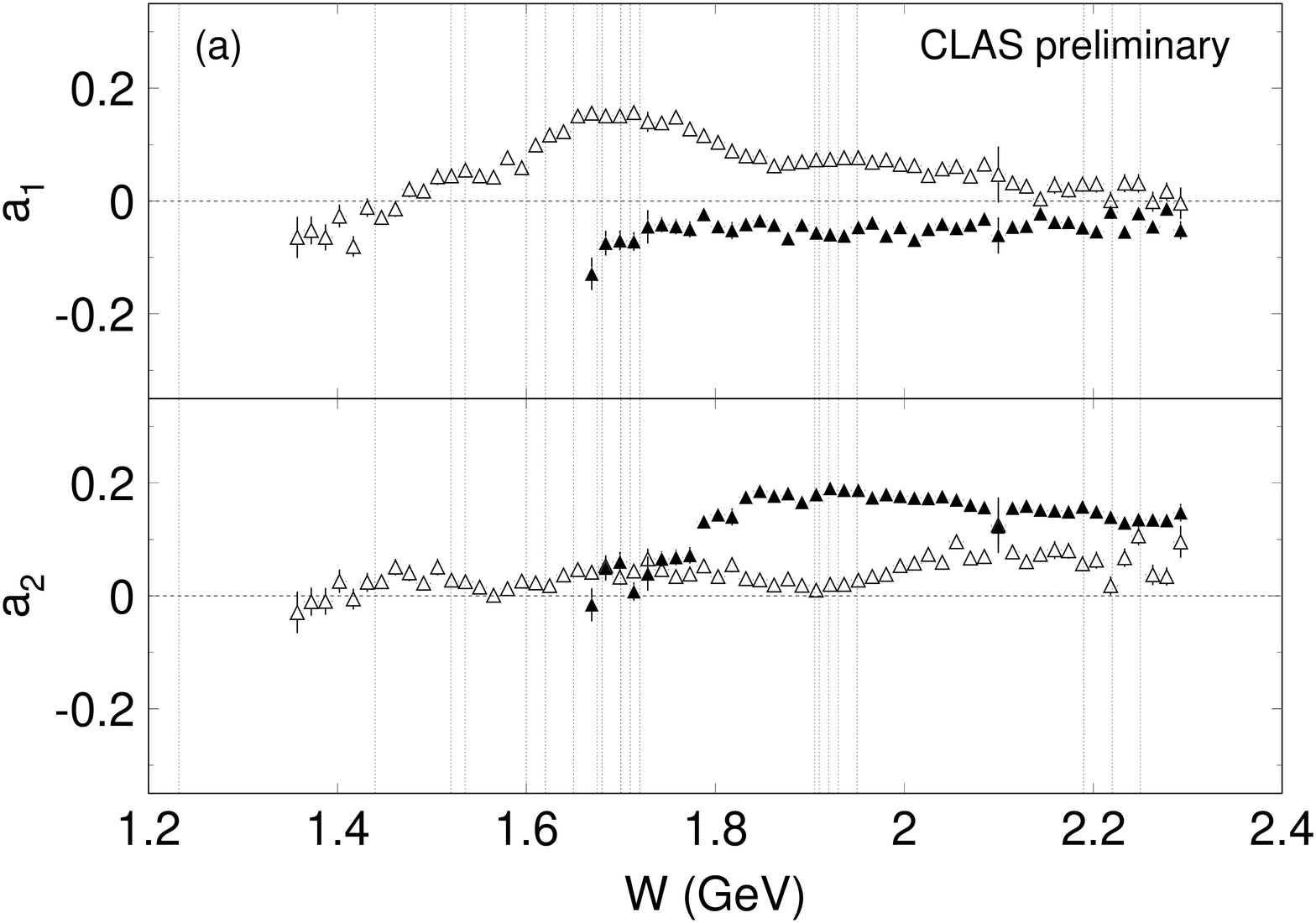,width=4.0in}}
\centerline{\epsfig{file=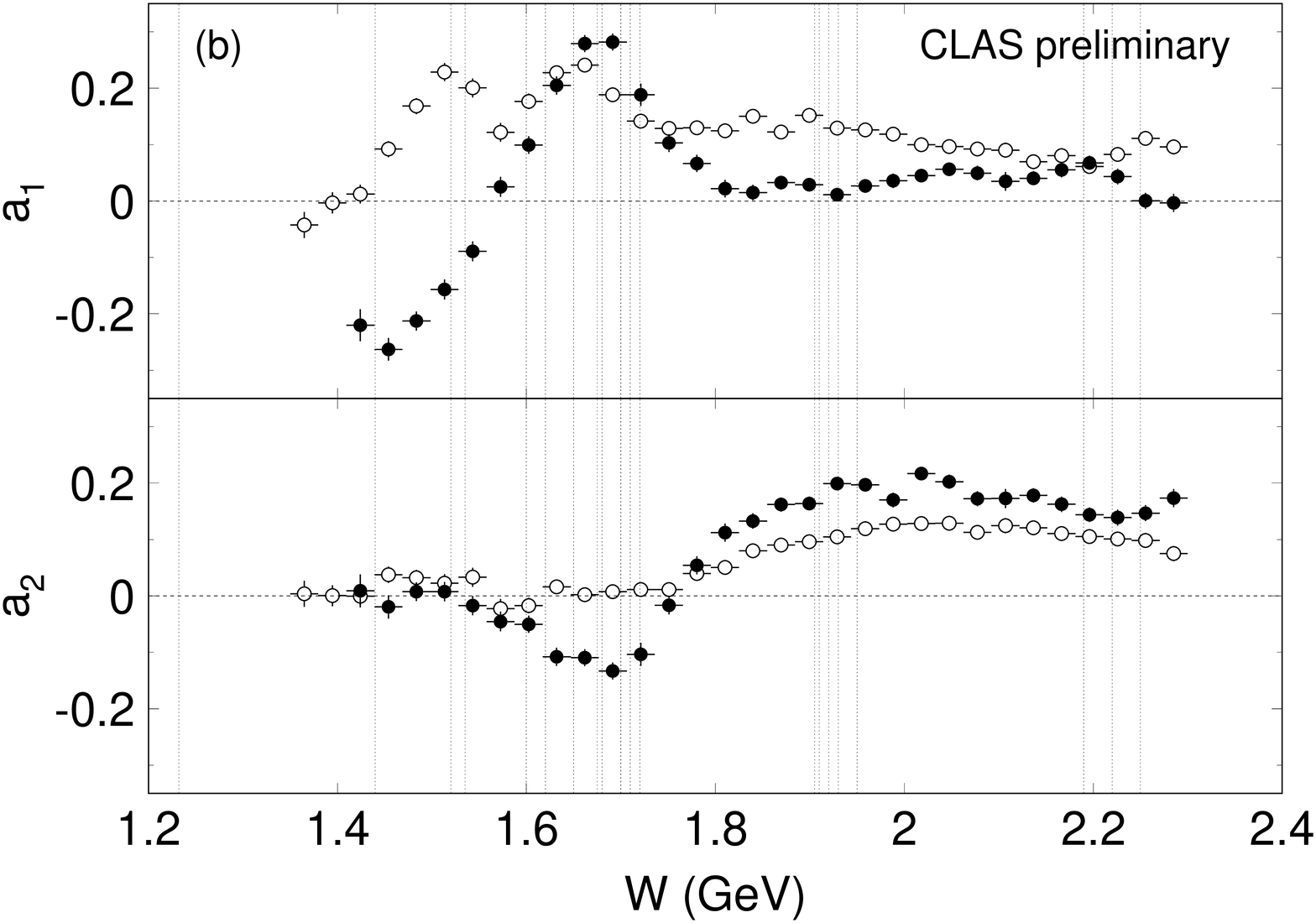,width=4.0in}}
\caption{Fourier components of the preliminary \gp cross-section
asymmetry as a function of the $\gamma N$ center-of-mass energy.  The
data were integrated over the full CLAS acceptance constrained by the
following cuts: panel (a) $\pi^+\pi^-$ invariant mass below 0.7~GeV
(open triangles) and above 0.7~GeV (filled triangles); panel (b)
$\cos(\Theta^*)$ less than $-0.30$ (open circles) and larger than
$+0.30$ (filled circles). The vertical lines indicate the masses of
3- and 4-star resonances \protect\cite{pdg02}.}
\label{fig:eg}
\end{figure}

The energy dependence of the $a_1$ and $a_2$ Fourier coefficients of
the observed preliminary angular distributions for the
$p[\pi^+,\pi^-]$ configuration is shown in Fig.~\ref{fig:eg}. To
demonstrate the strong dependence of the Fourier coefficients on the
photon energy, reaction channel, and details of the kinematics,
various cuts were applied to the data. Figure \ref{fig:eg}(a) shows
the data where the invariant mass of the $\pi^+\pi^-$ pair is less
than 0.7~GeV (open triangles) and more than 0.7~GeV (filled
triangles); Fig.~\ref{fig:eg}(b) shows different angular cuts, namely,
for data with $\cos(\Theta^*) \le -0.30$ (open circles) and for
$\cos(\Theta^*) \ge 0.30$ (filled circles). 

\begin{figure}[h!]
\centerline{\epsfig{file=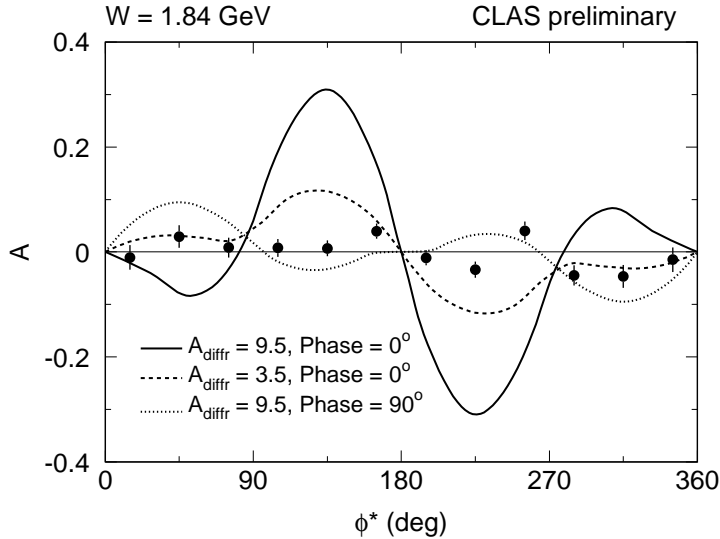,width=4.0in}}
\caption{Preliminary \gp angular distribution of the cross-section
asymmetry at $W = 1.84$ GeV for the $\pi^-[p,\pi^+]$ configuration,
along with calculations of Mokeev with various choices for
the diffractive rho-production amplitude \protect\cite{MokeevPC}.}
\label{fig:mokeev}
\end{figure}

It is desirable to examine the sensitivity of the asymmetry to various
theoretical model parameters. In fact, current studies have indicated
a strong sensitivity of the helicity asymmetries to relative
contributions of various isobaric channels and interference among
them.  An example is shown in Fig.~\ref{fig:mokeev}, where the
preliminary \gp asymmetry data at $W = 1.84$ GeV for the configuration
$\pi^-[p,\pi^+]$ are compared with various calculations by Mokeev
\cite{MokeevPC}. These calculations differ only in the amplitude and
phase of the diffractive rho-production term. The helicity asymmetry
shows strong sensitivity to this model parameter, whereas the
differential cross section does not. This confirms the importance of
the polarization observables for differentiating between model
predictions.

In summary, angular distributions of large helicity-dependent
cross-section asymmetries in the \gp reaction were observed for the
first time, using the CLAS detector system. Our preliminary analysis
has revealed the rich structure of these data, which are sensitive to
details of the reaction dynamics. These data will prove to be an
important tool in baryon spectroscopy, and similar data obtained using
light nuclei ($^2$H, $^3$He, and $^4$He) will help us to understand the
modification of baryonic parameters in the nuclear medium
\cite{Strauch03}.

This work was supported by the U.S.~Department of Energy under grant
DE--FG02--95ER40901. Southeastern Universities Research Association
(SURA) operates the Thomas Jefferson National Accelerator Facility
under U.S.~Department of Energy contract DE--AC05--84ER40150.


\begin{thebibliography}{99}
\bibitem{pdg02} K.~Hagiwara {\it et al.}, Phys.~Rev.~D {\bf 66}, 010001 (2002).
\bibitem{Ripani03} 
  M.~Ripani {\it et al.}, Phys.~Rev.~Lett. {\bf 91}, 022002 (2003).
\bibitem{Donnelly96} T.~W.~Donnelly, {\it Advances in Nuclear Physics,
  Vol. 22} (edited by J.~W.~Negele and E.~W.~Vogt, Plenum Press, New
  York, 1996), p.~37.
\bibitem{Boffi96} S.~Boffi, C.~Guisti, F.D.~Pacati, and M.~Radici,
  {\it Electromagnetic Response of Atomic Nuclei} (Clarendon Press,
  Oxford, 1996).
\bibitem{Mecking03}
  B.~A.~Mecking {\it et al.}, Nucl.~Instrum.~Methods {\bf A503}, 513
  (2003).
\bibitem{Sober00}
  D.~I.~Sober {\it et al.}, Nucl.~Instrum.~Methods {\bf A440}, 263 (2000).
\bibitem{Maximon59} 
  H.~Olsen and L.~C.~Maximon, Phys.~Rev.~{\bf 114}, 887 (1959).
\bibitem{Oed03} T.~Oed and W.~Roberts, private communication (2003).
\bibitem{Roberts97} W.~Roberts and A.~Rakotovao, hep-ph/9708236 for
  formalism.
\bibitem{Mokeev01} 
 V.~I.~Mokeev {\it et al.}, Phys.~At.~Nucl. {\bf 66}, 1282 (2003);
 V.~I.~Mokeev {\it et al.}, Phys.~At.~Nucl. {\bf 64}, 1292 (2001),
  and references therein.
\bibitem{MokeevPC} V.~I.~Mokeev, private communication (2003).
\bibitem{Strauch03} S.~Strauch {\it et al.}, Intl.~Conf.~on Few-Body
  Problems in Physics, Durham (2003).
\end{thebibliography}
\end{document}